\documentclass[10pt,conference]{IEEEtran}

\usepackage{amssymb}
\usepackage{amsxtra}
\usepackage{amsmath}
\usepackage{latexsym}
\usepackage{graphics}
\usepackage{verbatim}
\usepackage{epsfig}
\usepackage{setspace}
\usepackage{colortbl}
\usepackage{theorem}
\usepackage{algorithmic}
\usepackage{algorithm}

\theorembodyfont{\normalfont\slshape}

\newtheorem{theorem}{Theorem}[section]
\newtheorem{lemma}[theorem]{Lemma}
\newtheorem{proposition}[theorem]{Proposition}
\newtheorem{corollary}[theorem]{Corollary}

\newtheorem{remark}[theorem]{Remark}

\def\O{\Omega}
\def\S{\Sigma}

\def\Z{{\mathbb Z}}

\def\x{{\mathbf x}}

\def\cC{{\mathcal C}}

\def\cF{{\mathcal F}}
\def\cG{{\mathcal G}}
\def\cH{{\mathcal H}}
\def\cI{{\mathbf I}}

\def\cL{{\mathcal L}}

\def\cS{{\mathcal S}}
\def\cT{{\mathbf T}}

\def\cV{{\mathcal V}}

\def\cX{{\mathcal X}}
\def\cY{{\mathcal Y}}

\def\tr{{\text{tr}}}

\begin{document}
\thispagestyle{empty}

\title{The Zeta Function of a Periodic-Finite-Type Shift}

\author{Akiko Manada and Navin Kashyap\\
Department of Mathematics and Statistics\\
Queen's University, Kingston, ON, Canada, K7L3N6\\
\{akiko, nkashyap\}@mast.queensu.ca}

\maketitle

\renewcommand{\thefootnote}{} 
\footnotetext{This work was supported by a Discovery Grant from the 
Natural Sciences and Engineering Research Council (NSERC), Canada.}

\renewcommand{\thefootnote}{\arabic{footnote}}
\setcounter{footnote}{0}

\renewcommand{\markboth}[2]
{\renewcommand{\leftmark}{#1}\renewcommand{\rightmark}{#2}}

\markboth{}{}

\begin{abstract}
The class of periodic-finite-type shifts (PFT's) is a class of sofic shifts 
that strictly includes the class of shifts of finite type (SFT's),
and the zeta function of a PFT is a generating function for 
the number of periodic sequences in the shift. In this paper, 
we derive a useful formula for the zeta function of a PFT. 
This formula allows the zeta function of a PFT to be computed 
more efficiently than the specialization of a formula known 
for a generic sofic shift.
\end{abstract}

\section{Introduction\label{intro_sec}}

A sofic shift is a set of bi-infinite sequences which can be represented by 
some labeled directed graph, and is core to the study of constrained coding. 
A classic example of sofic shifts is the class of shifts of finite type 
(SFT's), which arise commonly in the context of coding for data storage.

A new class of sofic shifts, called \emph{periodic-finite-type shifts} 
(PFT's), was introduced by Moision and Siegel \cite{MS2},
who were interested in studying the properties of distance-enhancing codes,
in which the appearance of certain words is forbidden in a periodic manner. 
The class of PFT's strictly includes the class of SFT's, and
some other interesting classes of shifts, such as constrained systems
with unconstrained positions \cite{SMNW}, and shifts arising
from the time-varying maximum transition run constraint \cite{PM}.

The difference between the definitions of SFT's and PFT's is small,
but significant. An SFT is defined by forbidding the appearance 
of finitely many words at any position of a bi-infinite sequence. 
A PFT is also defined by forbidding the appearance of finitely many words
within a bi-infinite sequence, except that these words are only forbidden 
to appear at positions indexed by certain pre-defined \emph{periodic} 
integer sequences; see Section~\ref{basic_sec} for a formal definition. 
Thus, there is a notion of period inherent in the definition 
of a PFT that causes it to differ from an SFT.

The properties of SFT's are quite well understood 
(see, for example, \cite{LM}), but the same cannot be said for PFT's. 
The study of PFT's has, up to this point, primarily focused on finding 
efficient algorithms for constructing their presentations 
\cite{BCF},\cite{BCMS},\cite{CP}. The work presented in this 
paper may be viewed as part of an ongoing effort (see also \cite{MK}) 
to extend some of what is known about SFT's to the larger class of PFT's.

This paper focuses on zeta functions. The zeta function of 
a sofic shift is an exponential generating function for the 
number of periodic sequences in the shift. It is known that
the zeta function of a sofic shift is always a rational function \cite{M}.
The zeta function of a sofic shift can be explicitly computed 
from a labeled directed graph presenting the shift; 
see \cite[Theorem~6.4.8]{LM}. If the graph has $r$ vertices,
the formula requires the computation of the characteristic polynomials
of $r$ matrices. 

It is well known that the zeta function can be computed in a much
simpler way when the sofic shift is in fact an SFT.
In this case, the zeta function is obtainable from 
the characteristic polynomial of only one matrix --- 
the adjacency matrix of a graph derivable from 
a forbidden-word description of the SFT; see \cite[Theorem~6.4.6]{LM}. 
In this paper, we prove an analogous result 
(Theorem~\ref{zeta_function_thm}) for a PFT. 
We show that the zeta function of a PFT can be computed 
from certain matrices derivable from a description of the 
PFT in terms of periodically-forbidden words. 
Moreover, the number of these matrices depends only on the period
of the PFT. For example, the number of matrices needed is two
when the PFT has period equal to 2.

The rest of this paper is organized as follows. We provide some of the
necessary background on PFT's in Section~\ref{basic_sec}, and then move on 
to the derivation of the zeta function of a PFT in Section~\ref{zeta_sec}.

\section{Basic Background\label{basic_sec}}

We begin with the basic background, based on material from 
\cite{BCMS} and \cite{LM}. Let $\S$ be an \emph{alphabet}, 
a finite set of symbols. A \emph{word} over $\S$ is a finite-length
sequence $w$; the length of $w$ is denoted by $|w|$.
For a bi-infinite sequence 
$\x=\ldots x_{-1} x_{0} x_{1} \ldots $ over $\S$, 
we call a length-$n$ word $w$ a 
\emph{subword}\footnote{An alternative term for subword, used
in the literature on combinatorics on words, is ``factor'';
see e.g., \cite{lothaire}.}
of $\x$, denoted by $w \prec \x$, 
if $w=x_{i} x_{i+1} \ldots x_{i+n-1}$ for some integer $i$. 
We will write $w \prec _{i} \x$ when we want to emphasize the fact that 
$w$ is a subword of $\x$ starting at the index $i$. We further define
$\sigma^r(\x)=\ldots x^{*}_{-1} x^{*}_{0} x^{*}_{1} \ldots$, 
the \emph{r-shifted sequence} of $\x$, to be the bi-infinite sequence 
satisfying $x^{*}_{i}=x_{i+r}$ for all $i \in \Z$.     

Given a labeled directed graph $\cG$, where edge labels come from $\S$, 
let $S(\cG)$ be the set of bi-infinite sequences which are 
generated by reading off labels along bi-infinite paths in $\cG$. 
A \emph{sofic shift} $\cS$ is a set of bi-infinite sequences such that 
$\cS=S(\cG)$ for some labeled directed graph $\cG$. In this case, we say 
that $\cS$ is \emph{presented by} $\cG$, or that $\cG$ is a 
\emph{presentation} of $\cS$. A classic example of a sofic shift
is a \emph{shift of finite type} (\emph{SFT}) $\cY=\cY_{\cF'}$, 
where $\cF'$ is a finite set of forbidden words (a \emph{forbidden set}).
The SFT $\cY=\cY_{\cF'}$ is defined to be the set of all bi-infinite 
sequences $\x=\ldots x_{-1}x_{0}x_{1}\ldots$ over $\S$ such that 
$\x$ contains no word $f' \in \cF'$ as a subword. 

A \emph{periodic-finite-type shift} (\emph{PFT}) is characterized 
by an ordered list of finite sets 
$\cF=(\cF^{(0)},\cF^{(1)}, \ldots, \cF^{(T-1)})$ and a \emph{period} $T$. 
More precisely, the PFT $\cX_{\{\cF,T\}}$ is defined as 
the set of all bi-infinite sequences $\x$ over $\S$ 
such that for some integer $r\in \{0,1,\ldots, T-1\}$, 
the $r$-shifted sequence $\sigma^{r}(\x)$ of $\x$ satisfies 
$f \prec_i\sigma^{r}(\x)$ $\Longrightarrow $ 
$f \not \in \cF^{(i \mod T)}$ for every integer $i$. 
It is easy to see that a PFT $\cX_{\{\cF,T\}}$ with period $T=1$ is simply the 
SFT $\cY_{\cF'}$ with $\cF'=\cF^{(0)}$. 
Thus, the class of SFT's is (strictly) included in the class of PFT's.

Any PFT $\cX$ has a representation of the form $\cX_{\{\cF,T\}}$ 
such that $\cF^{(j)} = \emptyset$ for $1 \leq j \leq T-1$,
and every word in $\cF^{(0)}$ has the same length. 
An arbitrary representation $\cX_{\{\widehat \cF,T\}}$ 
can be converted to one in the above form as follows.
For a given PFT $\cX=\cX_{\{\widehat \cF,T\}}$, 
if $\hat f\in \widehat \cF^{(j)}$ for some $1\leq j\leq T-1$, 
then list out all words of length $j+|\hat f|$ which end with $\hat f$, 
add them to $\widehat \cF^{(0)}$, and delete $\hat f$ from 
$\widehat \cF^{(j)}$. Continue this process until 
$\widehat \cF^{(1)}=\cdots=\widehat \cF^{(T-1)}=\emptyset $. 
Next, find the longest word in the resulting $\widehat \cF^{(0)}$,
and let $\ell$ denote its length. Define $\cF^{(0)}=\{f\in \S^{\ell}: 
\mbox{$f$ starts with some word in $\widehat \cF^{(0)}$}\}$. 
It is easy to check that $\cX=\cX_{\{\cF,T\}}$ with 
$\cF=(\cF^{(0)}, \emptyset, \ldots, \emptyset)$,
and every word in $\cF^{(0)}$ has the same length $\ell$. 
Throughout this paper, for a given PFT $\cX=\cX_{\{\cF,T\}}$, 
we always assume that $\cF$ is in \emph{standard form}, 
\emph{i.e.}, $\cF=(\cF^{(0)}_{\cX},\emptyset, \ldots, \emptyset)$, 
and $\cF^{(0)}_{\cX}$ is a subset of $\S^{\ell}$ for some $\ell\geq 1$. 

Moision and Siegel proved that every PFT is a sofic shift, that is, 
each PFT has a presentation, by giving an algorithm to construct 
a presentation of a PFT \cite{BCMS}, \cite{MS2}. 
We call their algorithm the \emph{MS algorithm} and 
the resulting presentation $\cG_{\cX}$ of a PFT $\cX$
under the MS algorithm the \emph{MS presentation} of $\cX$. 
The MS algorithm, given a PFT $\cX=\cX_{\{\cF,T\}}$ 
with $\cF$ in standard form as input, runs as follows.\\[-18pt]

\begin{algorithm}[!h]
\caption {: The MS Algorithm}
\begin{algorithmic}[1]
\STATE define $T$ sets of words $\cV^{(0)}, \cV^{(1)}, \ldots, \cV^{(T-1)}$ 
as \\
$\cV^{(0)}=\S^{\ell}\setminus \cF^{(0)}_{\cX}$ 
and $\cV^{(1)}=\cV^{(2)}=\cdots=\cV^{(T-1)}=\S^{\ell}$.
\STATE \algorithmicfor\ each integer $0\leq j\leq T-1$,
{\bf and}\ \algorithmicfor\ each pair of words
$u=u_1u_2\ldots u_{\ell}\in \cV^{(j)}$ and 
$v=v_1v_2\ldots v_{\ell} \in \cV^{(j+1 \mod T)}$
\STATE \ \ \ \ \ \ \ \ \algorithmicif\ 
$u_2\ldots u_{\ell}=v_1\ldots v_{\ell-1}$ \algorithmicthen
\STATE \ \ \ \ \ \ \ \ \ \ \ \ \ \ draw an edge labeled 
$v_{\ell}$ from $u$ to $v$.
\STATE \ \ \ \ \ \ \ \ \algorithmicelse
\STATE \ \ \ \ \ \ \ \ \ \ \ \ \ \ draw no edge from $u$ to $v$.
\RETURN {the resulting directed graph and name it $\cG_{\cX}$.}
\end{algorithmic}
\end{algorithm}\mbox{}\\[-20pt]
We note here that B{\'e}al, Crochemore and Fici have also given an algorithm, 
different from the one above, that generates a presentation of a PFT 
\cite{BCF}. 

The MS presentation of a PFT is an example of a 
\emph{word-based graph} (\emph{WBG}), which 
we define to be any labeled digraph $\cG$ with the following properties:
\begin{itemize}
\item  every state in $\cG$ is a word $w\in \S^{\ell}$ for some $\ell\geq 1$.  
\item  the vertex set consists of $T$ disjoint \emph{phases} 
$\cV^{(0)}, \cV^{(1)}, \ldots, \cV^{(T-1)}$ for some $T\geq 1$, and
each phase has at most one state corresponding to $w\in \S^{\ell}$. 
We denote by $w^{(i)}$ the state in $\cV^{(i)}$ corresponding to $w$.
\item  there is an edge labeled $a\in \S$ from 
$u^{(i)}=u_1u_2 \ldots u_{\ell}\in \cV^{(i)} $ to 
$v^{(i+1 \mod T)}=v_1 v_2 \ldots v_{\ell} \in \cV^{(i+1 \mod T)}$  
if and only if $u_2\ldots u_{\ell}=v_1 \ldots v_{\ell-1}$ and $v_{\ell}=a$.  
\end{itemize} 
Observe that WBG's are always \emph{deterministic}, that is, 
distinct outgoing edges from the same state are labeled distinctly. 

Given a WBG $\cG$ and a path 
$\alpha:V_0\rightarrow V_1\rightarrow \cdots$ in $\cG$, 
let $\cI_{\cG}(\alpha)$ and $\cT_{\cG}(\alpha)$ denote 
the initial state and terminal state of $\alpha$ in $\cG$, respectively. 
In the case when $\cI_{\cG}(\alpha)=\cT_{\cG}(\alpha)=V$, 
we call $\alpha$ a \emph{cycle at $V$}. Furthermore, we denote by 
$\cL_{\cG}(\alpha)$ the sequence which is generated by reading off labels 
along $\alpha$. We also simply say $x$ is \emph{generated by} $\alpha$ 
(in $\cG$) if $x=\cL_{\cG}(\alpha)$. \\[-18pt]

\begin{remark}
For a WBG $\cG$ with $T$ phases,
\begin{itemize}
\item[(1)] if a path $\alpha:V_0\rightarrow V_1\rightarrow \cdots$ 
in $\cG$ satisfies $V_0\in \cV^{(i)}$, 
then $V_r\in \cV^{(k)}$ if and only if $r\equiv k-i \pmod T$; 
\item[(2)] 
there is no cycle of length $n$ in $\cG$ if $n\not \equiv 0 \pmod T$.
\end{itemize}
\label{cycle_remark}
\end{remark}

\section{The Zeta Function of a PFT\label{zeta_sec}}

The zeta function of a sofic shift $\cS$ is a generating function 
for the number of periodic sequences in $\cS$.
More precisely, for a given sofic shift $\cS$, 
the zeta function $\zeta _{\cS}(t)$ of $\cS$ is defined to be 
\begin{equation}
\zeta _{\cS}(t) =
  \exp\left(\sum_{n=1}^{\infty}\frac{|P_n(\cS)|}{n} \, t^n\right),
\label{zeta_def}
\end{equation}
where $P_n(\cS)$ is the set of periodic sequences in $\cS$ of period $n$. 
In fact, there exists a formula for computing the zeta function of a 
sofic shift, which shows that the zeta function of a sofic shift is 
always rational \cite{M},\cite[Theorem~6.4.8]{LM}.
In the particular case when $\cS$ is an SFT $\cY=\cY_{\cF'}$
with forbidden set $\cF'\subset \S^{\ell}$, the zeta function 
is simply the reciprocal of a polynomial \cite{BL}. 
More precisely (see \cite[Theorem~6.4.6]{LM}), 
$$\zeta _{\cY}(t)=\frac{1}{\det(I-tA_{\cG_{\cY}})},$$
where $A_{\cG_{\cY}}$ is the adjacency matrix of the 
MS presentation of $\cY$, by taking $\cY=\cX_{\{(\cF'),1\}}$. 

In this section, we derive an expression for the zeta function of 
a PFT $\cX_{\{\cF,T\}}$ that uses the adjacency matrices of certain WBG's
derivable from $\cF$ and $T$. This expression is given in 
Theorem~\ref{zeta_function_thm}.

We first note the following remark that can be easily verified from the 
properties of WBG's.\\[-18pt]

\begin{remark}
For any WBG $\cG$, a path $\alpha$ in $\cG$ of length $|\alpha|\geq \ell$ 
terminates at a state corresponding to 
$u=u_1u_2\ldots u_{\ell}\in \S^{\ell}$ 
if and only if the length-$\ell$ suffix of $\cL_{\cG}(\alpha)$ is $u$.
\label{terminal_state_rem} 
\end{remark}

We use Remark~\ref{terminal_state_rem} to prove 
Lemma~\ref{same_word_lemma} and Lemma~\ref{perseq_cycle_lemma}. 
which are important observations for proving 
Theorem~\ref{zeta_function_thm}.\\[-18pt]

\begin{lemma}
Let $\cG$ be a WBG, and
consider two states $u^{(i)}=u_1u_2\ldots u_{\ell}$ and 
$v^{(j)}=v_1v_2\ldots v_{\ell}$ (for some $i$ and $j$) in $\cG$. 
For two cycles $C$ and $\widetilde C$, $|C|=|\widetilde C|$, 
which are at $u^{(i)}$ and $v^{(j)}$, respectively,  
if $\cL_{\cG}(C)=\cL_{\cG}(\widetilde C)$, 
then both $u$ and $v$ are copies of the same word in $\S^{\ell}$.
\label{same_word_lemma}
\end{lemma} 
\emph{Proof\/}: 
Let $m$ be an integer such that $m|C|\geq \ell$. 
Since $(\cL_{\cG}(C))^{m}=(\cL_{\cG}(\widetilde C))^{m}$ 
and $|(\cL_{\cG}(C))^{m}|=m|C|\geq \ell$, 
both $\cT_{\cG}(C^m)$ and $\cT_{\cG}(\widetilde {C}^m)$ 
are the length-$\ell$ suffix of $(\cL_{\cG}(C))^{m}$ 
by Remark~\ref{terminal_state_rem}. Observe that 
$\cT_{\cG}(C^m)=\cT_{\cG}(C)=u^{(i)}$ and
$\cT_{\cG}(\widetilde {C}^m)=\cT_{\cG}(\widetilde C)=v^{(j)}$. 
\endproof\mbox{} \\[-20pt]

\begin{lemma}
Let $\cG_{\cX}$ be the MS presentation of a PFT 
$\cX=\cX_{\{\cF,T\}}$ with period $T$. 
When $n\equiv  0 \pmod T$, 
a periodic sequence $\x=(x_0x_1\ldots x_{n-1})^{\infty}$ 
is in $P_n(\cX)$ iff $x_0x_1\ldots x_{n-1}=\cL_{\cG_{\cX}}(C)$ 
for some cycle $C$ of length $n$ in $\cG_{\cX}$. 
\label{perseq_cycle_lemma}
\end{lemma}
\mbox{}\\[-4ex]
\emph{Proof\/}: 
If $x_0x_1\ldots x_{n-1}=\cL_{\cG_{\cX}}(C)$ for some cycle $C$ 
of length $n$ in $\cG_{\cX}$, then clearly  
$\x=(x_0x_1\ldots x_{n-1})^{\infty}$ is in $P_n(\cX)$ 
since $\x=(\cL_{\cG_{\cX}}(C))^{\infty}$.

Conversely, suppose that $\x=(x_0x_1\ldots x_{n-1})^{\infty}$ 
is in $P_n(\cX)$. Let $m$ be an integer such that $mn\geq \ell$. 
For a bi-infinite path $\alpha$ in $\cG_{\cX}$ satisfying 
$\cL_{\cG_{\cX}}(\alpha)=\x$, a finite subpath $\beta$ of $\alpha$
generating $(x_0x_1\ldots x_{n-1})^m$ terminates at 
$w^{(i)}$ for some $0\leq i\leq T-1$, where $w$ is the length-$\ell$ 
suffix of $(x_0x_1\ldots x_{n-1})^m$, by Remark~\ref{terminal_state_rem}.   
From $w^{(i)}$, there must be a path $\gamma$ generating 
$x_0x_1\ldots x_{n-1}$, and observe that 
$\cT_{\cG_{\cX}}(\gamma)=\cT_{\cG_{\cX}}(\beta\gamma)=w^{(j)}$ 
for some $0\leq j \leq T-1$. However, since $n\equiv 0  \pmod T$ 
by assumption, $i=j$ holds from the structure of the MS presentation 
$\cG_{\cX}$. It implies that there is a cycle (\emph{i.e.}, $\gamma$) 
at $w^{(i)}$ generating $x_0x_1\ldots x_{n-1}$. 
\endproof\mbox{} \\[-20pt]

Lemma~\ref{perseq_cycle_lemma} is not true when $n\not \equiv 0 \pmod T$ 
since, by (2) in Remark~\ref{cycle_remark}, there is no cycle of 
length $n$ in $\cG_{\cX}$. However, the following proposition shows
that we can generate $\x \in P_n(\cX)$ 
using a cycle $C$ in the MS presentation $\cG_{\cX_d}$ of $\cX_d$, 
where $\cX_d=\cX_{\{\cF_d,d\}}$ is the PFT with period $d=\gcd(n,T)$ and 
$\cF_d=(\cF^{(0)}_{\cX}, \emptyset ,\ldots,\emptyset )$.\\[-18pt]

\begin{proposition}
Let $\cX=\cX_{\{\cF,T\}}$ be a PFT with period $T$ and $\cF$ in standard form. 
If $\gcd(n,T)=d$, then $\x \in P_n(\cX)$ if and only if $\x \in P_n(\cX_d)$, 
where $\cX_d=\cX_{\{\cF_d,d\}}$ is the PFT with period $d$ and 
$\cF_d=(\cF^{(0)}_{\cX}, \emptyset ,\ldots,\emptyset )$.
\label{equiv_pro}
\end{proposition}
\mbox{}\\[-4ex]
\emph{Proof\/}: 
Clearly, $\x \in P_n(\cX_d)$ implies $\x \in P_n(\cX)$ by definition. 
So suppose that $\x \in P_n(\cX)$. 
If $\x \not\in P_n(\cX_d)$, then for any $0\leq r \leq d-1$, we have 
$f_r\prec _{\hat i_r}\sigma^{r}(\x)$ for some $f_r\in \cF^{(0)}_{\cX}$ 
and integer $\hat i_r\equiv 0 \pmod d$, that is, $f_r\prec _{i_r}\x$ 
for $i_r=\hat i_r +r\equiv r \pmod d$. 
Since $\gcd(n,T)=d$ by assumption, there exists an integer 
$m\geq 1$ such that $mn\equiv d \pmod T$.
As $\x$ is a periodic bi-infinite sequence of period $n$, 
for any $0\leq r \leq d-1$, $\x$ contains $f_r$ at indices 
$i_r+smn$, $s = 0,1,\ldots,T/d-1$. This implies that for each 
$r' \in \{0,1,\ldots, T-1\}$, we have $f_{r'}\prec _{j_r'}\sigma^{r'}(\x)$ 
for some $f_{r'} \in \cF^{(0)}_{\cX}$ and integer $j_{r'} \equiv 0 \pmod T$, 
which contradicts the fact that $\x\in P_n(\cX)\subset \cX$.     
\endproof\mbox{} \\[-20pt]

Thus, for the PFT's $\cX$ and $\cX_d$ defined in Proposition~\ref{equiv_pro}, 
when $\gcd(n,T)=d$, as $n\equiv 0 \pmod d$, 
we have from Lemma~\ref{perseq_cycle_lemma} that $\x \in P_n(\cX_d)=P_n(\cX)$ 
iff $\x=(\cL_{\cG_{\cX_d}}(C))^{\infty}$ for some cycle $C$ in the 
MS presentation $\cG_{\cX_d}$ of $\cX_d$.  
That is, for any $n$, every periodic sequence $\x \in P_n(\cX)$ 
can be generated by a cycle in the MS presentation of some PFT. 
       
So, for a WBG $\cG$, let $\cC_{n}(w^{(i)})_{\cG}$ be 
the set of periodic sequences $\x=(x_0x_1\ldots x_{n-1})^{\infty}$
which can be generated by a cycle $C$ of length $n$ at $w^{(i)}$ in $\cG$. 
In other words, $\x=(x_0x_1\ldots x_{n-1})^{\infty} 
\in \cC_{n}(w^{(i)})_{\cG}$ iff there exists a cycle $C$ 
at $w^{(i)}$ satisfying $\cL_{\cG}(C)=x_0x_1 \ldots x_{n-1}$. 
Putting together Lemma~\ref{same_word_lemma}, 
Lemma~\ref{perseq_cycle_lemma} and Proposition~\ref{equiv_pro}, 
we have the following corollary.\\[-18pt]

\begin{corollary}
Let $\cX=\cX_{\{\cF,T\}}$ be a PFT with period $T$ and $\cF$ in standard form. 
Given an integer $n \geq 1$, suppose $\gcd(n,T) = d$, and consider the PFT 
$\cX_d=\cX_{\{\cF_d,d\}}$ with period $d$ and 
$\cF_d=(\cF^{(0)}_{\cX}, \emptyset ,\ldots,\emptyset )$.
Then, we have that
\begin{equation}
|P_n(\cX)|=|P_n(\cX_d)|=\sum_{w\in \S^{\ell}}
  \left|\bigcup _{i=0}^{d-1}\cC_{n}(w^{(i)})_{\cG_{\cX_d}}\right|.
\label{num_per_seq_equ}
\end{equation}
\mbox{}\\[-4ex]
\label{sum_cor}
\end{corollary} 
\mbox{}\\[-4ex]
\emph{Proof\/}: 
The first equality is obvious from Proposition~\ref{equiv_pro}. 
Also, observe from Lemma~\ref{perseq_cycle_lemma} that 
$|P_n(\cX_d)|$ is equal to $|\bigcup_{w\in \S^{\ell}}
\bigcup _{i=0}^{d-1}\cC_{n}(w^{(i)})_{\cG_{\cX_d}}|$, and 
$\cC_{n}(w^{(i)})_{\cG_{\cX_d}}\cap \cC_{n}
(\widehat {w}^{(j)})_{\cG_{\cX_d}}=\emptyset$ 
if $w\not=\widehat w$ by Lemma~\ref{same_word_lemma}, 
which shows the second equality.
\endproof\mbox{} \\[-20pt]

For (\ref{num_per_seq_equ}), we have from the inclusion-exclusion principle
that 
$$
\left|\bigcup_{i=0}^{d-1}\cC_{n}(w^{(i)})_{\cG_{\cX_d}}\right|=
\sum_{J \subseteq [d]}(-1)^{|J|-1}
\left|\bigcap_{j\in J} \cC_n(w^{(j)})_{\cG_{\cX_d}}\right|,
$$
where $[d]=\{0,1,\ldots, d-1\}$. 
Therefore, our goal is to count $|\bigcap_{j\in J} \cC_n(w^{(j)})_{\cG_{\cX_d}}|$ 
for each $J\subset [d]$. To do this, we define WBG's determined by 
certain binary words, as discussed next.  
 
We need a few definitions. For a word $z$ over some alphabet,
let $z^\sharp$ denote the \emph{primitive root} of $z$, 
\emph{i.e.}, $z^\sharp$ is the shortest word such that 
$z = {(z^\sharp)}^{n}$ for some integer $n \geq 1$. 
In particular, for a binary word $z$ (over the alphabet $\{0,1\}$), 
we will find it convenient to define $L_z$ to be the length of $z^\sharp$, 
$W_z$ to be the number of $1$'s in $z^\sharp$, and $N_z = |z|/L_z$.
Thus, $z = {(z^\sharp)}^{N_z}$.

Now, let $\cX = \cX_{\{\cF,T\}}$ be a PFT with period $T$ and $\cF$ in
standard form. Set $B_T = \{0,1\}^T \setminus \{0^T\}$. 
For $z =z_0 z_1 \ldots z_{T-1} \in B_T$, 
let $\sigma_T(z)$ denote the cyclically shifted sequence
$z_{1} z_{2} \ldots z_{T-1} z_0$. We say that $z,z' \in B_T$
are \emph{conjugate} (or \emph{cyclically equivalent}) 
if $z' = \sigma_T^q(z)$ for some integer $q$. Conjugacy is 
an equivalence relation on $B_T$, which partitions $B_T$
into \emph{conjugacy classes}. We then construct a set $\O_T$ 
by picking from each conjugacy class of $B_T$ 
one representative sequence $z = z_0 z_1 \ldots z_{T-1}$ such that $z_0 = 1$.

For each $z = z_0z_1\ldots z_{T-1} \in \O_T$, 
let us denote by $\cG_z$ the word-based graph with $L_z$ phases 
$\cV^{(0)}, \cV^{(1)}, \ldots, \cV^{(L_z-1)}$ defined 
(for $i = 0,1,\ldots,L_z-1$) by 
$$
\cV^{(i)} =
\begin{cases}
  \S^{\ell} & \mbox{when}\ z_i = 0.\\  
  \S^{\ell}\setminus \cF^{(0)}_{\cX}  & \mbox{when}\ z_i = 1.
\end{cases}
$$ 
For example, $\cG_{10^{T-1}}$ is the MS presentation $\cG_{\cX}$ of $\cX$.
Furthermore, let us denote by $\cH_{z}$ the subgraph of $(\cG_{z})^{L_z}$ 
induced by $\cV^{(0)}=\Sigma^{\ell}\setminus \cF^{(0)}_{\cX}$. 
More precisely, the vertex set of $\cH_{z}$ is $\cV^{(0)}$ in $\cG_{z}$, and 
there is an edge from $u^{(0)}$ to $v^{(0)}$ in $\cH_{z}$ iff 
there is a path of length $L_z$ from $u^{(0)}$ to $v^{(0)}$ in $\cG_{z}$.
We denote by $A_z$ and $B_z$ the adjacency matrices of 
$\cG_z$ and $\cH_{z}$, respectively. 

The following lemma states an important relationship between
the traces of the matrices $A_z$ and $B_z$ defined above. \\[-18pt]

\begin{lemma}
(a)\ $\tr(A_z^n) = 0$ if $n \not\equiv 0 \pmod{L_z}$. \\
(b)\ $\tr(A_z^{L_z m})=L_z \times \tr(B_z^m)$ for any integer $m \geq 1$. 
\label{trace_lemma} 
\end{lemma}
\mbox{}\\[-4ex]
\emph{Proof\/}: (a) follows directly from Remark~\ref{cycle_remark}(2).
For (b), first observe that for any integer $m\geq 1$,
$\tr(A_z^{L_z m})=\sum_{i=0}^{L_z-1}\tr(B_{\cV^{(i)}}^m)$, 
where $B_{\cV^{(i)}}$ is the adjacency matrix of the subgraph 
of $(\cG_{z})^{L_z}$ induced by $\cV^{(i)}$.
However, we also have $\tr(B_{\cV^{(0)}}^m)=\tr(B_{\cV^{(1)}}^m)
= \cdots = \tr(B_{\cV^{(L_z-1)}}^m)$,
since every cycle of length $L_zm$ in $\cG_z$ can be viewed
as a cycle at a state in $\cV^{(i)}$ for any $0\leq i\leq L_z-1$.
\endproof\mbox{} \\[-20pt]

We clarify at this point that our reason for defining 
the matrices $A_z$, $B_z$ is that they are an integral part of 
our zeta function expression in Theorem~\ref{zeta_function_thm}.
Precisely how these matrices enter into that expression will become clearer
after the next couple of lemmas. The first of these lemmas
explains the relationship between states in 
the MS presentation $\cG_{\cX}$ of a PFT $\cX$ 
and a state in the graph $\cG_{z}$ defined from a $z \in \O_T$.\\[-18pt]

\begin{lemma}
Let $\cX=\cX_{\{\cF,T\}}$ be a PFT with period $T$ and $\cF$ in standard form.
For each $z = z_0z_1 \ldots z_{T-1} \in \O_T$ 
and each integer $0 \leq q \leq L_{z}-1$, define
\begin{equation}
J(z,q) = \{(q - i) \ \text{mod}\,\, T: 0 \leq i \leq T-1, \, z_i = 1\}
\label{J_eq1}
\end{equation}
The mapping $(z,q) \mapsto J(z,q)$ is a bijection between pairs $(z,q)$
as above and non-empty subsets $J \subseteq [T]$. Furthermore, 
setting $J = J(z,q)$, we have
\begin{equation} 
{\cC_n(w^{(q)})}_{\cG_{z}}=\bigcap_{j\in J} {\cC_n(w^{(j)})}_{\cG_{\cX}},
\label{wbg_equiv_equ} 
\end{equation}
\mbox{}\\[-2ex]
for any integer $n \equiv 0 \pmod T$, and any word $w \in \S^{\ell}$.
\label{count_lem}
\end{lemma} 
\mbox{}\\[-4ex]
\emph{Proof\/}:
It is easy to verify that the mapping $(z,q) \mapsto J(z,q)$ is a bijection
as stated above, so we focus on proving (\ref{wbg_equiv_equ}).
%
%
For clarity, we use the notation 
$\cV^{(i)}_{\cG_{\cX}}$ and $\cV^{(i)}_{\cG_{z}}$ (for some $i$)
to denote the phase $\cV^{(i)}$ in $\cG_{\cX}$ and the phase $\cV^{(i)}$ 
in $\cG_{z}$, respectively. 

Consider $\x=(x_0x_1\ldots x_{n-1})^{\infty}\in \cC_n(w^{(q)})_{\cG_{z}}$. 
Then there exists a cycle 
$C:V_0=w^{(q)}\rightarrow V_1\rightarrow \cdots \rightarrow V_{n-1}=w^{(q)}$ 
of length $n$ at $w^{(q)}\in \cV^{(q)}_{\cG_z}$ such that 
$\cL_{\cG_z}(C)=x_0 x_1 \ldots x_{n-1}$. 
Recall from Remark~\ref{cycle_remark}(1) that $V_r\in \cV^{(i')}_{\cG_z}$, 
where $0\leq i' \leq L_z-1$, iff $r\equiv i'-q\pmod {L_{z}}$. 
Since $\cV^{(i')}_{\cG_z}=\S^{\ell}\setminus \cF^{(0)}_{\cX}$ 
iff $z_{i'}=1$ for some $0\leq i' \leq L_z-1$,
$V_r$ cannot be a word in $\cF^{(0)}_{\cX}$ iff 
$r\equiv i'-q\pmod {L_{z}}$ for some $0\leq i'\leq L_z-1$ 
satisfying $z_{i'}=1$. Furthermore, as 
$z_{i'}=z_{i'+L_z}=\cdots=z_{i'+(N_z-1)L_z}$ for each $0\leq i' \leq L_z-1$, 
we infer that $V_r$ cannot be a word in $\cF^{(0)}_{\cX}$ iff 
$r\equiv i-q\pmod {T}$ for some $0\leq i\leq T-1$ satisfying $z_i=1$. 

Now, consider $\x'=(x'_0x'_1\ldots x'_{n-1})^{\infty}\in \bigcap_{j\in J} 
\cC_n(w^{(j)})_{\cG_{\cX}}$ for $J=\{j_1,j_2, \ldots, j_{|J|}\}=J(z,q)$. 
Then, for each $j \in J$ and a state $w^{(j)}\in \cV^{(j)}_{\cG_{\cX}}$,
there exists a cycle 
$C^{(j)}:w^{(j)}\rightarrow V_1^{(j)} \rightarrow \cdots\rightarrow  
V_{n-2}^{(j)}\rightarrow w^{(j)}$ of length $n$ 
at $w^{(j)}\in \cV^{(j)}_{\cG_{\cX}}$ such that 
$\cL_{\cG_{\cX}}(C^{(j)})=x'_0x'_1\ldots x'_{n-1}$. 
For the cycle $C^{(j)}$, we have, from Remark~\ref{cycle_remark}(1), 
that $V_r^{(j)} \in \cV^{(0)}_{\cG_{\cX}}$ iff 
$r\equiv -j \pmod T$. That is, $V_r^{(j)}$ cannot be a word 
in $\cF^{(0)}_{\cX}$ iff  $r \equiv -j \pmod T$. 
Since $\cL_{\cG_{\cX}}(C^{(j_1)})=\cL_{\cG_{\cX}}(C^{(j_2)})
=\cdots=\cL_{\cG_{\cX}}(C^{(j_{|J|})})$, 
we have that for each cycle $C^{(j)}$, $j \in J$, 
$V^{(j)}_r$ cannot be a word in $\cF^{(0)}_{\cX}$ \
iff  $r \equiv -j_k \pmod T$ for some $j_k\in J$.
Since $J=J(z,q)$ is as defined in (\ref{J_eq1}), we find that 
$V^{(j)}_r$ cannot be a word in $\cF^{(0)}_{\cX}$ iff 
$r\equiv i-q \pmod {T}$ for some $0\leq i\leq T-1$ satisfying $z_i=1$.

Hence, $x_0x_1\ldots x_{n-1}=\cL_{\cG_z}(C)$  
for some cycle $C$ at $w^{(q)}\in \cV^{(q)}_{\cG_z}$ 
iff for any $j\in J$, there exists a cycle $C^{(j)}$ 
at $ w^{(j)}\in \cV^{(j)}_{\cG_{\cX}}$ such that 
$x_0x_1\ldots x_{n-1}=\cL_{\cG_\cX}(C^{(j)})$. This clearly shows that 
$(z,q)$ and $J=J(z,q)$ satisfy (\ref{wbg_equiv_equ})  
for any integer $n\equiv 0 \pmod T$ and any $w\in \S^{\ell}$,
as required. 
\endproof\mbox{}\\[-20pt]

We are now in a position to give the key idea in the proof
of our zeta function result. As we show in the next lemma,
we can explicitly determine $|P_n(\cX)|$ for a PFT $\cX=\cX_{\{\cF,T\}}$ 
using the adjacency matrices of the WBG's $\cG_z$. 

\begin{lemma}
Let $\cX=\cX_{\{\cF,T\}}$ be a PFT with period $T$ and $\cF$ in standard form. 
For an integer $n\geq 1$, suppose that $\gcd(n,T)=d$ and consider 
the PFT $\cX_d=\cX_{\{\cF_d,d\}}$ with period $d$ and 
$\cF_d=(\cF^{(0)}_{\cX}, \emptyset ,\ldots,\emptyset )$. Then, 
$$
 |P_n(\cX)| = |P_n(\cX_d)|
   = \sum_{z\in \O_d}(-1)^{d W_z/L_z - 1}
	\mbox{tr} (A_z^n).
$$
\label{trace_lem}
\end{lemma}
\mbox{}\\[-2ex]
\emph{Proof\/}: 
We would like to show, from Corollary~\ref{sum_cor}, that 
\begin{eqnarray*}
\lefteqn{\sum_{w\in \S^{\ell}}
         \sum_{J \subseteq [d]}(-1)^{|J|-1}
         \left|\bigcap_{j\in J} \cC_n(w^{(j)})_{\cG_{\cX_d}}\right|} \\
& & = \ \ \sum_{z\in \O_d}(-1)^{d W_z/L_z - 1}
	\mbox{tr} (A_z^n).
\end{eqnarray*}
\mbox{}\\[-2ex]
Pick $w\in \S^{\ell}$ arbitrarily. From Lemma~\ref{count_lem},
we see that the mapping that takes a pair $(z,q)$, with
$z \in \O_d$ and $0 \leq q \leq L_z-1$, to
\begin{equation}
J(z,q) = \{(q - i) \ \text{mod}\,\, d: 0 \leq i \leq d-1, \, z_i = 1\}
\label{J_eq2}
\end{equation}
gives us a one-to-one correspondence between such pairs $(z,q)$,
and non-empty sets $J \subseteq [d]$, satisfying (\ref{wbg_equiv_equ}).
Furthermore, as WBG's are deterministic, 
we have that for any state $w^{(q)}$ in $\cG_z$, 
$|\cC_n(w^{(q)})_{\cG_{z}}|$ is equal to the 
number of cycles of length $n$ at $w^{(q)}$ in $\cG_z$, 
which is the $(w^{(q)}, w^{(q)})$-th entry of $A_z^n$. 
Finally, it is clear from (\ref{J_eq2}) that 
$|J(z,q)|$ is equal to the number of $1$'s in $z$, 
which in turn is equal to $(d/L_z)W_z$. These observations 
are enough to prove the lemma.
\endproof\mbox{}\\[-2ex]

We use the above lemma to derive our expression for the zeta function
of a PFT $\cX = \cX_{\{\cF,T\}}$. We first re-write the sum 
 $\sum_{n=1}^\infty \frac{|P_n(\cX)|}{n} t^n$ in (\ref{zeta_def}) as
\begin{equation}
\sum_{d|T} \sum_{n:\gcd(n,T)=d} \frac{|P_n(\cX)|}{n} \, t^n.
\label{sum1}
\end{equation}
Using Lemma~\ref{trace_lem}, the sum above can be expressed as 
\begin{equation}
\sum_{d|T} 
  \sum_{z\in \O_d}
   \sum_{n:\gcd(n,T)=d} (-1)^{d W_z/L_z-1} 
	\frac{\mbox{tr} \left(A_z^n\right)}{n} \, t^n.
\label{sum2}
\end{equation}
Observe from the definition of $\cG_z$ that for $z\in \O_T$,
$\cG_z=\cG_{\hat z}$ iff $\hat z$ can be represented 
as $\hat z=(z^\sharp)^k$, for some positive integer $k$. 
Therefore, for $z\in \O_T$, $A_z$ appears in (\ref{sum2}) iff
$d=kL_z$ for some $k|N_z$. Thus, (\ref{sum2}) can be expressed as 
\begin{equation}
\sum_{z\in \O_T}
 \sum_{k|N_z} 
   \sum_{n:\gcd(n,T)=kL_z} (-1)^{k W_z-1} 
	\frac{\mbox{tr} \left(A_z^n\right)}{n} \, t^n.
\label{sum2'}
\end{equation}
A standard application of the M{\"o}bius inversion formula of 
elementary number theory allows us to write (\ref{sum2'}) as
$$
\sum_{z\in \O_T}
  \sum_{k|N_z} 
  \sum_{r|\frac{N_z}{k}} \mu(r) 
   \sum_{m=1}^\infty (-1)^{k W_z-1} 
   \frac{\mbox{tr} \left( A_z^{rkL_zm}\right)}{rkL_zm} \, t^{rkL_zm},
$$
where $\mu(\cdot)$ is the M\"{o}bius function. Using the change of variable 
$s = rk$ (so that the sum over pairs $(k,r)$ is now a sum over pairs $(s,r)$),
the above may be rewritten as
\begin{eqnarray}
\lefteqn{\!\!\!\!\!\!\!\!\! 
\sum_{z \in \O_T} \sum_{s | N_z} \sum_{r | s} \mu(r) (-1)^{(s/r)W_z - 1}
\sum_{m=1}^\infty \frac{\tr(A_z^{s L_z m})}{s L_z m} \, t^{s L_z m}} \notag \\
&=& \sum_{z \in \O_T} \sum_{s | N_z} \beta(z,s)
\sum_{m=1}^\infty \frac{\tr(A_z^{s L_z m})}{s L_z m} \, t^{s L_z m},
\label{sum4}
\end{eqnarray}
where we have defined $\beta(z,s) = \sum_{r | s} \mu(r) (-1)^{(s/r)W_z - 1}$.
A simple expression for $\beta(z,s)$, stated in the next lemma,
can be obtained using the standard M{\"o}bius function fact 
that, for any positive integer $k$, 
we have $\sum_{r|k} \mu(r) = 1$ if $k = 1$, 
and $\sum_{r|k} \mu(r) = 0$ if $k > 1$
A complete proof of the lemma will be published in the full version 
of this paper. 

\begin{lemma}
For any $z\in \O_T$ and $s|N_z$, we have 
$$
\beta(z,s) 
= 
\begin{cases}
(-1)^{W_z-1} & \text{ if } s = 1 \\
-1 - (-1)^{W_z - 1} & \text{ if } s = 2 \\
0 & \text{ otherwise}
\end{cases}
$$
\label{beta_lemma}
\end{lemma}

We are now in a position to prove our main result, namely, 
a formula for the zeta function of a PFT. \\[-18pt] 

\begin{theorem}
For a PFT $\cX = \cX_{\{\cF,T\}}$ with period $T$ and $\cF$ in standard form, 
the zeta function $\zeta_{\cX}(t)$ of $\cX$ is given by 
\begin{eqnarray*}
\zeta_\cX(t) &=& 
   \prod_{z \in \O_T} {[\det(I - t A_z)]}^{(-1)^{W_z}} \\ 
& & \times \prod_{\stackrel{\mbox{\scriptsize $z \in \O_T:$}}
                    {N_z \text{ is even }, \ W_z \text{ is odd }}}
     \det(I - t^{2L_{z}} {B_z}^2).
\end{eqnarray*}
\label{zeta_function_thm}
\end{theorem}
\mbox{}\\[-4ex]
\emph{Proof\/}: 
We first simplify the expression in (\ref{sum4}) using Lemma~\ref{beta_lemma}.
For $z$'s such that $N_z$ is odd or $W_z$ is even 
(in this case, by the lemma, $\beta(z,2) = 0$ as well), we observe that 
\begin{eqnarray*}
\lefteqn{\sum_{s | N_z} \beta(z,s)
\sum_{m=1}^\infty \frac{\tr(A_z^{s L_z m})}{s L_z m} \, t^{s L_z m}} \\
&=& 
(-1)^{W_z - 1} \sum_{m=1}^\infty \frac{\tr(A_z^{L_z m})}{L_z m} \, t^{L_z m} \\
&=& (-1)^{W_z - 1} \sum_{n=1}^\infty \frac{\tr(A_z^n)}{n} \, t^{n},
\end{eqnarray*}
as $\tr(A_z^n)=0$ if $n\not \equiv 0 \pmod {L_z}$, by Lemma~\ref{trace_lemma}.
Similarly, for $z$'s such that $N_z$ is even and $W_z$ is odd, we have
\begin{eqnarray*}
\lefteqn{\sum_{s | N_z} \beta(z,s)
\sum_{m=1}^\infty \frac{\tr(A_z^{s L_z m})}{s L_z m} \, t^{s L_z m}} \\ 
&=& 
(-1)^{W_z - 1} \sum_{m=1}^\infty \frac{\tr(A_z^{L_z m})}{L_z m} \, t^{L_z m}\\ 
& &\ \; \; 
+ \ (-2) \, \sum_{m=1}^\infty \frac{\tr(A_z^{2L_z m})}{2L_z m} \, t^{2L_z m} \\
&=& (-1)^{W_z - 1} \sum_{n=1}^\infty \frac{\tr(A_z^n)}{n} \, t^{n} 
- \sum_{m=1}^\infty \frac{\tr(B_z^{2m})}{m} \, t^{2L_{z}m},
\end{eqnarray*}
where the last equality comes from Lemma~\ref{trace_lemma}. 

The theorem now follows by plugging these expressions into (\ref{sum4}), 
and then using the fact (see e.g., \cite[Theorem~6.4.6]{LM})
that for any square matrix $A$ and positive integer $k\geq 1$,
\begin{equation}
\exp\Big(\sum_{n=1}^{\infty}\frac{\mbox{tr}(A^{kn})}{n}t^{kn}\Big)
={[\det(I-t^kA^k)]}^{-1}. \notag
\end{equation}
\mbox{}\\[-20pt]
\endproof\mbox{}\\[-20pt]

Theorem~\ref{zeta_function_thm} clearly shows that  
the zeta function of a PFT is rational, and 
the number of graphs (or adjacency matrices) needed to compute 
the zeta function of a PFT depends only on its period $T$. 
Furthermore, the case when $N_z$ is even and $W_z$ is odd can happen 
only when the period $T$ is even. Therefore, when $T$ is odd, 
we can compute the zeta function using only the adjacency matrices $A_z$:
$$
\zeta_\cX(t) = \prod_{z \in \O_T} {[\det(I - t A_z)]}^{(-1)^{W_z}}.
$$


\begin{thebibliography}{99}
\bibitem{BCF}
M.-P.\ B{\'e}al, M.\ Crochemore and G.\ Fici, 
``Presentations of constrained systems with unconstrained positions,''
\emph{IEEE Trans.\ Inform.\ Theory},
vol.\ 51, no.\ 5, pp.\ 1891--1900, May 2005.


\bibitem{BL}
R.\ Bowen and O.E.\ Lanford, 
``Zeta functions of restrictions of the shift transformation,''
\emph{Proc.\ Symp.\ Pure Math.\ A.M.S.},
vol.\ 14, pp.\ 43--50, 1970.

\bibitem{BCMS}
M.-P.\ B{\'e}al, M.\ Crochemore, B.E.\ Moision and P.H.\ Siegel,
``Periodic finite-type shift spaces,'' preprint 
submitted to \emph{IEEE Trans.\ Inform.\ Theory}.

\bibitem{CP} D.P.B.\ Chaves and C.\ Pimentel,
``An algorithm for finding the Shannon cover of 
a periodic shift of finite type,'' preprint
submitted to \emph{IEEE Trans.\ Inform.\ Theory}.

\bibitem{LM} D.\ Lind and B.H.\ Marcus,
\emph{An Introduction to Symbolic Dynamics and Coding},
Cambridge University Press, 1995. 

\bibitem{lothaire} M.\ Lothaire, \emph{Combinatorics on Words},
Cambridge Univ.\ Press, 1997.

\bibitem{MK} A.\ Manada and N.\ Kashyap, 
``On the periods of a periodic-finite-type Shift,'' 
\emph{Proc.\ 2008 IEEE Int.\ Symp.\ Inform.\ Theory (ISIT 2008)},
 Toronto, Canada, pp.\ 1453--1457, July 2008.

\bibitem{M} A.\ Manning, 
``Axiom A diffeomorphisms have rational zeta functions,'' 
\emph{Bull. London Math. Soc.}, vol.\ 3, pp.\ 215--220, 1971.


\bibitem{MS2} B.E.\ Moision and P.H.\ Siegel, 
``Periodic-finite-type shift spaces,'' 
\emph{Proc.\ 2001 IEEE Int.\ Symp.\ Inform.\ Theory (ISIT'01)},
Washington DC, June 24--29, 2001, p.\ 65.

\bibitem{PM} T.L.\ Poo and B.H.\ Marcus,
``Time-varying maximum transition run constraints,''
\emph{IEEE Trans.\ Inform.\ Theory},
vol.\ 52, no.\ 10, pp.\ 4464--4480, Oct.\ 2006.

\bibitem{SMNW}
J.C.\ de Souza, B.H.\ Marcus, R.\ New and B.A.\ Wilson, 
``Constrained systems with unconstrained positions,''
\emph{IEEE Trans.\ Inform.\ Theory},
vol.\ 48, no.\ 4, pp.\ 866--879, April 2002.

\end{thebibliography}
\end{document}